\documentclass{iau}
\usepackage{graphicx}

\title[Abundances in Stars with Debris Disks] 
{Abundances in Stars with Debris Disks}

\author[A. M. Ritchey \etal\ ]   
{Adam M. Ritchey$^1$, Guillermo Gonzalez$^2$, Myra Stone$^1$, 
\and George Wallerstein$^1$}

\affiliation{$^1$Department of Astronomy, University of Washington, \\ Box 
351580, Seattle, WA 98195, USA \\ email: {\tt aritchey@astro.washington.edu} 
\\[\affilskip]
$^2$Department of Physics and Astronomy, Ball State University, \\ 2000 W. University Ave., Muncie, IN 47306, USA }

\pubyear{2013}
\volume{299}  
\pagerange{?--?}
\jname{Exploring the Formation and Evolution of Planetary Systems}
\editors{B. Matthews \& J. Graham, eds.}
\begin{document}

\maketitle

\begin{abstract}
We present preliminary results of a detailed chemical abundance analysis for a sample of solar-type stars known to exhibit excess infrared emission associated with dusty debris disks. Our sample of 28 stars was selected based on results from the Formation and Evolution of Planetary Systems (FEPS) \emph{Spitzer} Legacy Program, for the purpose of investigating whether the stellar atmospheres have been polluted with planetary material, which could indicate that the metallicity enhancement in stars with planets is due to metal-rich infall in the later stages of star and planet formation. The preliminary results presented here consist of precise abundances for 15 elements (C, O, Na, Mg, Al, Si, S, Ca, Sc, Ti, V, Cr, Fe, Co, and Ni) for half of the stars in our sample. We find that none of the stars investigated so far exhibit the expected trend of increasing elemental abundance with increasing condensation temperature, which would result from the stars having accreted planetary debris. Rather, the slopes of linear least-squares fits to the abundance data are either negative or consistent with zero. In both cases, our results may indicate that, like the Sun, the debris disk host stars are deficient in refractory elements, a possible signature of terrestrial and/or gas giant planet formation.

\end{abstract}

\firstsection 
\section{Sample Selection and Instrumentation}
We initially selected 23 stars from those identified by \cite[Hillenbrand \etal\ (2008)]{h08} as having ``Tier 1'' or ``Tier 2'' debris disks that had declinations above $-30^{\circ}$ and spectral types between F5 and K5. To this list, we added 12 additional stars with infrared excesses from \cite[Carpenter \etal\ (2009)]{c09}. Each star was observed with the echelle spectrograph of the 3.5~m telescope at Apache Point Observatory, which provides nearly complete optical wavelength coverage at a resolving power of $R\approx31,500$. The exposure times were designed to yield signal-to-noise ratios of approximately 300:1. After reducing the data, we found that 7 stars displayed broad, blended absorption features, making them unsuitable for detailed chemical abundance analyses. Thus, these stars were dropped from the sample, leaving a total of 28 stars. Here, we present detailed results for 14 of the 28.

\section{Detailed Chemical Abundance Analysis}
Basic stellar parameters were determined spectroscopically through the usual excitation and ionization analysis and abundances were derived for 15 elements using the latest version of the LTE spectral analysis code MOOG. Our line list includes 53 Fe I lines, 6 Fe II lines, and 50 lines from other atomic species carefully chosen to be free from blends and telluric interference. Results of our spectroscopic analysis for the initial 14 stars are given in Table~\ref{tab1} along with the dust parameters listed in \cite[Hillenbrand \etal\ (2008)]{h08}.

\begin{table}
\begin{center}
\caption{Derived Stellar Parameters, Slopes, and Dust Parameters from \cite[Hillenbrand \etal\ ]{h08}}
\label{tab1}
\begin{tabular}{lcccccc}\hline
Star & $T_{\mathrm{eff}}$ (K) & log $g$ & [Fe/H] & Slope$^1$ & $T_{\mathrm{dust}}$ (K) & $R_{\mathrm{dust}}$ (AU) \\
\hline
HD~377 & $5701\pm57$\phantom{0} & $4.01\pm0.06$ & $-0.01\pm0.05$ & $-25.31\pm3.43$ & 58 & 23 \\
HD~6963 & $5422\pm53$\phantom{0} & $4.23\pm0.07$ & $-0.19\pm0.04$ & $-20.08\pm3.84$ & 56 & 18 \\
HD~8907 & $6301\pm62$\phantom{0} & $4.37\pm0.06$ & $+0.00\pm0.05$ & $-10.42\pm3.37$ & 48 & 49 \\
HD~17925 & $5068\pm55$\phantom{0} & $4.13\pm0.11$ & $+0.06\pm0.03$ & $-12.92\pm5.67$ & 110 & 4 \\
HD~25457 & $6260\pm120$ & $4.1\pm0.2$ & $+0.11\pm0.09$ & \phantom{0}$-8.66\pm5.62$ & 70 & 23 \\
HD~70573 & $5761\pm46$\phantom{0} & $4.30\pm0.08$ & $-0.05\pm0.03$ & \phantom{0}$-4.69\pm2.81$ & 41 & 35 \\
HD~104860 & $5987\pm94$\phantom{0} & $4.35\pm0.08$ & $+0.07\pm0.07$ & \phantom{0}$-8.03\pm3.32$ & 46 & 42 \\
HD~107146 & $5848\pm62$\phantom{0} & $4.40\pm0.06$ & $+0.00\pm0.05$ & \phantom{0}$-7.44\pm2.93$ & 52 & 30 \\
HD~122652 & $6208\pm42$\phantom{0} & $4.46\pm0.07$ & $+0.12\pm0.03$ & \phantom{0}$-9.02\pm2.06$ & 55 & 31 \\
HD~145229 & $5796\pm57$\phantom{0} & $4.02\pm0.13$ & $-0.18\pm0.04$ & $-13.97\pm3.55$ & 54 & 26 \\
HD~150706 & $5857\pm49$\phantom{0} & $4.46\pm0.06$ & $-0.05\pm0.04$ & $-13.88\pm3.01$ & 58 & 23 \\
HD~201219 & $5537\pm51$\phantom{0} & $4.33\pm0.07$ & $+0.09\pm0.04$ & \phantom{0}$-3.21\pm3.66$ & 53 & 23 \\
HD~204277 & $6373\pm88$\phantom{0} & $4.6\pm0.1$ & $+0.11\pm0.06$ & \phantom{0}$-1.80\pm3.53$ & $<$50 & 43 \\
HD~206374 & $5520\pm44$\phantom{0} & $4.40\pm0.08$ & $-0.11\pm0.03$ & $-11.92\pm2.89$ & $<$74 & 12 \\
\hline
\end{tabular}
\end{center}
\vspace{1mm}
\scriptsize{
{\it Note:} $^1$Slope of weighted linear least-squares fit to [X/H] versus $T_{\mathrm{cond}}$ (in units of 10$^{-5}$ dex K$^{-1}$).}
\end{table}

\section{Summary and Conclusions}
We have completed a detailed chemical abundance analysis for 14 of the 28 debris disk stars in our sample, deriving basic stellar parameters and abundances for 15 elements.\footnote{Complete results for the full sample will be presented in Ritchey \emph{et al.} (in preparation).} All stars examined so far exhibit flat or negative slopes in plots of abundance ([X/H]) versus condensation temperature ($T_{\mathrm{cond}}$; taken from \cite[Lodders 2003]{l03}), suggesting that the stellar atmospheres have not accreted refractory-rich planetary material. Since the Sun has been found to be deficient in refractory elements relative to solar twins (\cite[Mel\'endez \etal\ 2009]{m09}), a result interpreted as a signature of terrestrial planet formation, and since these stars have compositions very similar to the Sun (and in some cases show even more of a deficiency in refractories), our results may indicate that planet formation has indeed occurred in these systems, and that the refractory elements are locked up in terrestrial planets or giant planet cores.

\section{Future Work}
Beyond completing the analysis for the remaining 14 stars in our sample, we will add abundance measurements for Zn since the derived slopes are heavily dependent on the abundances of the few elements with relatively low condensation temperatures (i.e., C, O, and S). We will also examine the impact of NLTE effects on the C and O abundances for the stars in our sample because, again, these effects could influence the slopes in some non-negligible way.


\end{document}